\documentclass{article}

\usepackage{PRIMEarxiv}
\usepackage[utf8]{inputenc}
\usepackage[T1]{fontenc}
\usepackage[hidelinks]{hyperref}
\usepackage{url}
\usepackage{booktabs}
\usepackage{amsfonts}
\usepackage{nicefrac}
\usepackage{microtype}
\usepackage{fancyhdr}
\usepackage{graphicx}
\graphicspath{{media/}}
\usepackage{listings}
\usepackage{xcolor}
\usepackage{multirow}
\usepackage{array}
\usepackage{amsmath}
\usepackage{float}

\lstdefinestyle{solidity}{
    basicstyle=\ttfamily\small,
    breaklines=true,
    frame=single,
    keywordstyle=\color{blue},
    commentstyle=\color{gray},
    stringstyle=\color{red},
    showstringspaces=false,
    tabsize=4,
}
\title{Mitigating Error Propagation in Chain-of-Thought: A Tree-of-Thought Framework for Smart Contract Repair}

\author{
  \begin{tabular}[t]{c}
    Jingping Zhu \\
    Hainan University \\
    Haikou, China \\
    \texttt{15937191601@163.com}
  \end{tabular}
  \hfill
  \begin{tabular}[t]{c}
    Hongping Wang \\
    Hainan University \\
    Haikou, China \\
    \texttt{hobinary.wong@gmail.com}
  \end{tabular}
  \hfill
  \begin{tabular}[t]{c}
    Xiaoqi Li \\
    Hainan University \\
    Haikou, China \\
    \texttt{csxqli@ieee.org}
  \end{tabular}
}

\begin{document}
\maketitle

\begin{abstract}
Smart contracts power blockchain applications such as DeFi and NFTs. However, once deployed, they cannot be modified. Even minor bugs can result in significant financial losses. Current AI-based repair methods rely on linear reasoning, which leads to the accumulation of errors and unreliable patches. Our method combines document parsing, static analysis, and Tree of Thoughts reasoning. We first convert audit reports into structured data. Then we use Slither to locate the exact vulnerable code. Our three-step framework explores multiple repair paths simultaneously, evaluates options, and eliminates poor choices. Finally, we verify patches through compilation and manual checks. We test our method on 50 real vulnerabilities from Code4Rena. Our method achieves a 62\% single success rate and an 84\% top-3 success rate, outperforming ContractTinker by 12 and 6 percentage points, respectively. We also increase the proportion of fully effective patches to 44\%, while reducing defective patches from 38\% to 22\% and invalid patches from 10\% to 4\%. This approach overcomes the limitations of linear reasoning and makes smart contract repair more accurate and practical.
\end{abstract}

\section{Introduction}

Blockchain technology has transformed traditional business practices in fields such as finance, supply chain management, digital assets, government services, and intelligent transportation~\cite{zhu2024sybil}. Smart contracts are a core application of blockchain, they significantly reduce transaction costs, improve execution efficiency, and enable trusted transactions without the need for third parties. Since Ethereum's launch, over 50 million smart contracts have been deployed on the Ethereum mainnet alone, providing technical support for the circulation of digital assets and applications such as DeFi, NFTs, gambling applications, and the metaverse~\cite{werner2022sok, long2025fomo3d}. However, this scale of growth has also brought greater security risks~\cite{zhou2023sok, 10172700, gao2025implementation}. Once deployed, smart contracts are difficult to modify, leaving vulnerabilities exposed in the public domain for extended periods. Attackers who discover vulnerabilities can exploit them repeatedly, causing cascading losses~\cite{gudgeon2020decentralized, Ding2025ACS}. In 2016, the infamous The DAO attack exploited a reentrancy vulnerability to steal approximately \$60 million worth of Ether, directly leading to a hard fork of the Ethereum blockchain. In 2023, the Curve Finance protocol was attacked due to an integer overflow vulnerability, resulting in losses exceeding \$70 million. In recent years, numerous techniques for identifying smart contract vulnerabilities have emerged~\cite{tolmach2021survey, kalra2018zeus, tsankov2018securify, he2019learning, zhuang2020smart, Hu2023LargeLM}. While these methods are effective at pinpointing issues~\cite{durieux2020empirical}, vulnerability fixes primarily rely on manual effort. Developers must analyze the root causes of vulnerabilities, assess their impact, and then write and verify patch code. When projects undergo frequent iterations or contract logic is complex, this process can significantly prolong the time required for fixes. Additionally, some developers lack security experience and may introduce new defects during the repair process~\cite{chaliasos2024smart}.

In recent years, advancements in large language models and multi-agent technologies have provided new approaches to the automated repair of smart contract vulnerabilities. The powerful capabilities of large language models in code understanding and generation allow them to learn vulnerability patterns and repair methods from vast amounts of code and security data, they have already demonstrated practical utility in program repair tasks. However, existing large language model-based methods for repairing smart contract vulnerabilities are still not fully mature. Models may suffer from hallucination, generating patches that appear reasonable but are actually incorrect. When domain knowledge of smart contracts is lacking, patches may deviate from business requirements. Linear, single-path reasoning makes results highly sensitive to prompts and context, leading to significant fluctuations in patch quality. At the same time, many methods lack rigorous patch verification steps, making it difficult to promptly detect cases where a fix ``appears to resolve the issue but actually introduces new risks.'' In contrast, traditional template or search-driven methods are more stable but have limited coverage, low generation efficiency, and insufficient capability to handle vulnerabilities involving complex control flows or cross-function dependencies. Therefore, combining the reasoning capabilities of large language models with formal methods such as static code analysis to improve accuracy and verifiability while ensuring efficiency is a key technical challenge that needs to be addressed in the current field of smart contract automatic repair. This paper focuses on this issue and proposes a thought-tree-based method for the automatic repair of smart contract vulnerabilities. By generating and filtering multiple path candidates, this method mitigates the problem of inconsistent patch quality, providing a more reliable automated solution for smart contract security. It addresses issues in existing smart contract vulnerability auto-repair methods, such as significant fluctuations in patch quality, low repair success rates, and the lack of a verifiable patch validation process. This paper proposes an auto-repair method that combines structured document parsing, static program analysis, and multi-branch reasoning based on thought trees.

This paper makes the following contributions:

(1) We propose and implement a module for the structured parsing of audit reports. To address the inconsistencies in report formats across different audit agencies, this module employs multi-pattern regular expression matching and semantic segmentation to extract key information. From the text, it converts natural language reports into computable structured data and thereby reduces the comprehension costs associated with format heterogeneity.

(2) We build a Slither-based static analysis module. This module parses contract code to generate a function call graph, extracting program semantic information. Simultaneously, through multi-granularity matching, it maps vulnerability descriptions in natural language to specific code entities, providing large language models with more focused and accurate context, thereby reducing window pressure and interference from irrelevant information caused by directly inputting the entire contract.

(3) We design a vulnerability repair reasoning framework based on a thought tree to mitigate the problem of errors amplifying at each stage in a linear thought chain.
\section{Related Work}

\subsection{Program Repair}

As an emerging research direction in the field of blockchain security~\cite{zhou2025blockchain, bobadilla2025automated, Zhang2025DoSAA}, the automatic repair of smart contract vulnerabilities has seen extensive research in recent years, with researchers exploring various technical approaches. This has led to the formation of four major technical frameworks: template-based repair, search-based repair, large language model-based repair, and generative-verification-based repair.

Template-based repair is one of the earliest methods for automatically fixing smart contract vulnerabilities. This approach is relatively straightforward, it first prepares fixed patch templates for a specific class of vulnerabilities, then uses static analysis tools to locate the exact position of the vulnerability, and finally inserts the template into the corresponding location to complete the fix. In 2021, Nguyen et al.\ proposed the sGuard template-based repair tool~\cite{nguyen2021sguard}, which combines symbolic execution and dependency analysis to identify four common types of vulnerabilities: reentrancy, integer overflow, and abuse of `tx.origin', among others. sGuard primarily injects patches through standardized rewriting. For example, it inserts functions from secure math libraries, adds the `nonReentrant' modifier, or replaces `tx.origin' with `msg.sender'~\cite{gao2024sguardplus}. Experimental results show that this tool can fix 65.4\% of target vulnerabilities, with an introduced gas overhead of approximately 0.79\%. EVMPatch~\cite{263790}, proposed by Rodler et al., performs templated fixes at the bytecode level. It applies online patches to deployed contracts by inserting jump code into proxy contracts. Developers do not need to implement complex upgradeable architectures~\cite{li2026uscsa} themselves and can complete upgrades in a short time, thereby circumventing the practical limitation that ``on-chain contracts cannot be modified.'' However, the limitations of the template-based patching approach are also evident. Its coverage depends on a predefined set of templates. It is often ineffective against complex business logic errors and novel vulnerabilities~\cite{10172700, bobadilla2025automated}. To accommodate general scenarios, templates often introduce additional instructions and state checks, which may lead to an unintended increase in gas costs. More troubling still, some seemingly ``safe'' fixes can alter control flow or permission boundaries. Under edge cases, these modifications may break the original business semantics.

Search-based repair methods model vulnerability repair as an optimization problem: finding the highest-scoring modification within a predefined space of patches. They typically apply mutations to the original code to generate a large number of candidate patches, which are then filtered one by one using test cases or formal verification tools, retaining only those that pass the checks. The SCRepair tool proposed by Yu et al.~\cite{doi_10_1145_3402450} is a representative example of this class of methods. This tool employs a genetic algorithm as its search strategy, defines three mutation operators (shift, replace, and insert), and introduces a gas advantage relationship to prioritize patches with lower gas consumption. Experimental results show that the tool can fix 54.2\% of the target vulnerabilities, and the generated patches reduce average gas overhead by 9.31\%. However, search-based repair methods are prone to search space explosion. As the scale of a smart contract increases, the generation and verification of candidate patches significantly increase computational overhead and time consumption. A more practical issue is that these methods heavily rely on high-quality test cases to determine whether a patch is correct. However, in engineering practice, it is difficult to prepare a test suite that covers all business logic. Consequently, even if a patch passes existing tests, it may still introduce functional regressions.

Generative-verification-based fixes combine the flexibility of generative methods with the constraints of formal verification. The system first iteratively generates candidate patches and then uses a verifier to check the correctness of each patch one by one. This allows the set of candidate patches to be narrowed down to a range that is provably safe. SmartFix, proposed by So et al.~\cite{doi_10_1145_3611643_3616341}, is a typical implementation of this approach. SmartFix adopts a generate-verify architecture that uses a statistical model to incorporate feedback from the verifier, the system adjusts the patch search order to prioritize modifications that are more likely to pass verification. Experimental results show that SmartFix achieves a 94.8\% success rate in fixing five common vulnerability categories, including integer overflow and re-entry. This figure is higher than sGuard's 65.4\%. At the same time, the patches generated by SmartFix are shorter. It introduces only 26.9\% of the number of runtime checks used by sGuard.

With the development of large language models, the automatic repair of smart contract vulnerabilities using these models has become a key research focus in recent years~\cite{fan2023automated, xia2023keep, sobania2023analysis, prenner2022can, Li2026DefensibleDF, jiang2023impact, yang2022vulrepair, tufano2022empirical, jimenez2024swebench, Huang2023AgentCoderMC}. Research indicates that real-world vulnerabilities are often not low-level syntax errors but rather high-level issues closely related to business logic~\cite{wu2025exploring}. Methods based on large language models leverage their code understanding and generation capabilities to directly generate patches by combining vulnerability descriptions with contract context, without relying on predefined repair templates. Consequently, they are better equipped to handle vulnerabilities with complex logic and variable types. The LLM-BSCVM framework proposed by Jin et al.~\cite{doi_10_1007_978_981_95_8417_8_4} is a representative work in this research direction. This framework introduces multi-agent collaboration and combines retrieval-enhanced generation, breaking down the repair process into six subtasks (including detection, analysis, repair, and evaluation, etc), which are carried out by different agents. Experimental results show that the framework achieves vulnerability detection accuracy and F1 scores both exceeding 91\% on benchmark datasets, while reducing the false positive rate to 5.1\%. SCPatcher~\cite{li2026scpatcher} similarly leverages large language models, combining retrieval-augmented generation with a knowledge graph to automate smart contract code repair. Addressing the challenge that traditional automated repair tools struggle to detect vulnerabilities in real business logic, ContractTinker employs a combination of ``large language models + linear thought chains + static analysis.'' It focuses on high-level vulnerabilities involving business logic issues and achieves better repair performance than pattern-based methods on real-world high-risk vulnerability datasets. However, such methods generally rely on linear thought chain reasoning, once an early judgment deviates, subsequent steps often continue along the incorrect path, causing errors to be progressively amplified~\cite{doi_10_1145_3691620_3695349}.

Although existing research on the automatic repair of smart contract vulnerabilities has made significant progress, there are still notable shortcomings at the methodological level. Template-driven approaches rely on predefined rules and can cover only a limited range of vulnerability types. Search-based repair often requires iterative trial-and-error within a large patch space, resulting in high time overhead. The generate-verify paradigm shifts the burden of ensuring correctness to extensive verification, causing computational costs to rise rapidly as the number of candidate patches increases. Most existing methods based on large language models employ single-path linear reasoning, which cannot effectively handle complex vulnerability repair tasks, making it difficult to guarantee the correctness and stability of patches. Therefore, how to fully leverage the large language models' ability to understand complex business logic to build an efficient, accurate, and secure framework for the automatic repair of smart contract vulnerabilities is a critical issue that urgently needs to be addressed in this field.

\subsection{Reasoning Chains in Large Language Models}

\begin{figure}[htbp]
  \centering
  \includegraphics[width=0.8\linewidth]{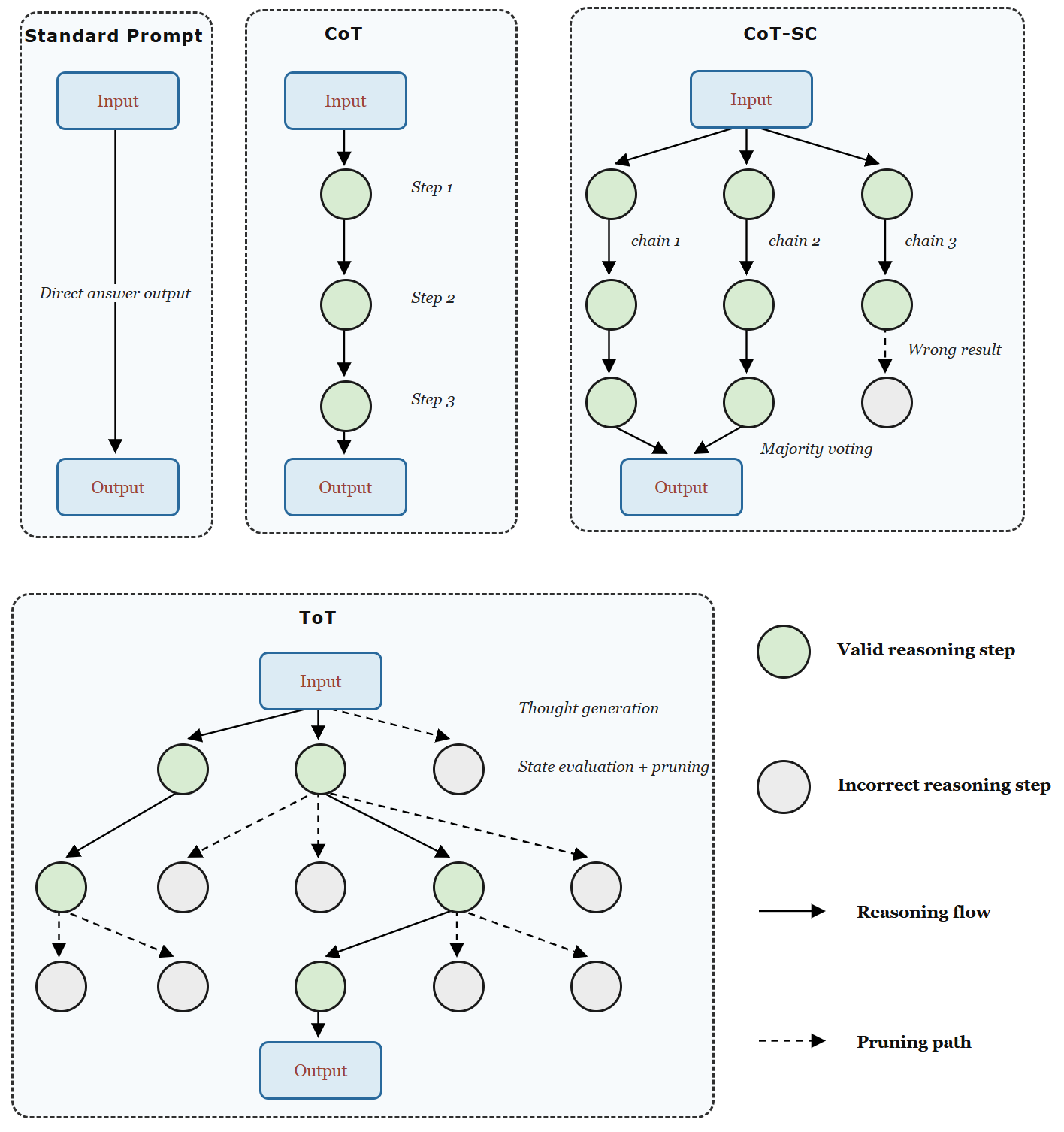}
  \caption{Comparison of the reasoning processes for standard prompts, CoT, CoT-SC, and ToT.}
  \label{fig:fig1}
\end{figure}

Large-model reasoning technology has evolved through the basic input-output prompt paradigm, the thought chain reasoning paradigm, the self-consistent thought chain reasoning paradigm, and the thought tree reasoning paradigm, gradually transitioning from single-path linear reasoning to multi-path structured reasoning. This paper focuses on the three classic paradigms mentioned above for further discussion. Figure~\ref{fig:fig1} compares the reasoning processes of standard prompts, Chain of Thought (CoT), Self-Consistent Chain of Thought (CoT-SC), and Tree of Thought (ToT). Green nodes represent valid reasoning steps, white nodes represent erroneous steps, solid lines indicate the flow of reasoning, and dashed lines indicate pruned, invalid paths.

\subsubsection{Chain of Thought}

CoT is a landmark technology in the field of large language model reasoning. First proposed by Wei et al.~\cite{3600270.3602070}, it fundamentally transforms the traditional paradigm of prompt engineering, which focused solely on direct input-output mappings. Prior to the emergence of CoT, large language models perform poorly on tasks requiring multi-step reasoning. Models often provide answers directly, skipping intermediate reasoning steps, which leads to high error rates. The implementation of CoT is very straightforward, simply include the phrase ``Let's think through this step by step'' in the prompt~\cite{3600270.3601883}, or demonstrate the complete reasoning process in an example. This method leverages the language understanding and generation capabilities of large language models to break down complex problems into several smaller ones, solving them one by one to ultimately arrive at the correct answer. Research shows that CoT technology works best on sufficiently large language models, when the model size reaches a certain threshold, its reasoning ability experiences a significant leap. However, the standard CoT method generates only a single reasoning path, if an error occurs at any intermediate step, the subsequent reasoning process will deviate from the correct direction, ultimately leading to an incorrect answer. Furthermore, CoT cannot evaluate or correct the generated reasoning steps, nor can it backtrack to previous steps to reconsider them, which limits its performance when handling complex problems that require exploring multiple possibilities~\cite{nye2022show, suzgun2023challenging, wang2023plan, zhou2023least, zhang2023automatic}.

\subsubsection{Self-Consistent Chain of Thought}

To overcome the error-prone nature of the single reasoning path in standard CoT, Wang et al.~\cite{wang2023iclr-selfconsistency} proposed the CoT-SC technique. The core idea of this technique is to leverage the sampling diversity of large language models to generate multiple independent reasoning paths and corresponding answers, and then select the answer with the highest frequency through majority voting as the final result. The working principle of CoT-SC is based on the observation that, for a given problem, the correct answer can often be derived through multiple different reasoning paths, whereas incorrect answers typically correspond to only a few reasoning paths. By generating multiple reasoning paths and counting the frequency of answers, it is possible to effectively filter out erroneous answers that occur by chance, thereby improving the robustness of the reasoning results. Compared to standard thought chains, self-consistent thought chains achieve better performance on nearly all tasks requiring reasoning, with particularly significant improvements in mathematical and logical reasoning tasks. The implementation process of CoT-SC primarily consists of three steps: Random sampling methods are used to have the large language model generate multiple different reasoning paths and answers. Then, all generated answers are analyzed to calculate the frequency of each answer. Finally, the answer with the highest occurrence frequency is selected as the final result. It should be noted that the number of generated reasoning paths requires a trade-off between performance and computational cost; typically, generating 5-20 reasoning paths yields good results. Although CoT-SC significantly improves reasoning performance, it still has some shortcomings. Generating multiple reasoning paths increases computational cost, which can become particularly high when handling complex problems. When different reasoning paths yield different answers with similar occurrence frequencies, the majority voting method may fail to select the correct answer. Furthermore, self-consistent reasoning chains still cannot evaluate the quality of reasoning steps nor proactively correct erroneous ones~\cite{fu2023complexity, madaan2023selfrefine}.

\subsubsection{Tree of Thoughts}

ToT is a more advanced reasoning paradigm for large language models proposed by Yao et al.~\cite{NEURIPS2023_271db992}. It models the problem-solving process as a tree-like structure, allowing the model to systematically explore, evaluate, and backtrack, thereby enabling it to solve complex problems that require deep reasoning and multi-step planning. Compared to CoT and CoT-SC, ToT more closely resembles the way humans solve problems. When solving complex problems, humans typically try various approaches, evaluate the feasibility of each, and, when encountering a dead end, backtrack to a previous step to reconsider their options. ToT's core framework consists of four components: thought decomposition, thought generation, state evaluation, and search algorithms. The thought decomposition phase breaks down complex problems into a series of ordered thinking steps. Each step corresponds to a node in the tree. The thought generation phase generates multiple possible ideas for each node, these ideas represent different approaches to solving the problem. The state evaluation phase assesses the quality of each idea to determine whether it has the potential to ultimately solve the problem. The search algorithm phase selects promising ideas based on the evaluation results to continue exploring. It prunes lower-quality ideas and, when necessary, returns to a previous node to start over. ToT supports multiple search algorithms, such as breadth-first search, depth-first search, and Monte Carlo tree search. Different search algorithms are suited for different types of problems. Breadth-first search is suitable for problems requiring the exploration of multiple possibilities. Depth-first search is suitable for problems requiring in-depth exploration of a single line of reasoning. By flexibly selecting search algorithms, the thought tree can adapt to a wide variety of reasoning tasks. ToT performs exceptionally well on many tasks requiring deep reasoning. Compared to CoT and CoT-SC, ToT can explore the problem space more systematically, avoiding getting stuck in local optima. However, ToT's implementation is also more complex, requiring the design of appropriate thought decomposition methods, thought generation strategies, and state evaluation functions. Additionally, its computational cost is higher because it must generate and evaluate ideas across multiple nodes~\cite{yao2023react, shinn2023reflexion, besta2024graph, hao2023reasoning, long2023large}.
\section{Theoretical Analysis}

This section addresses the shortcomings of the ContractTinker method mentioned earlier, which were exposed in real-world vulnerability repair scenarios. Using a real high-risk vulnerability in the MarginSwap project as a case study, we identify the core limitations of the CoT method. Building on this, we introduce the ToT method, design a ToT module tailored to smart contract vulnerability repair scenarios, and optimize the ContractTinker method to resolve the core issues of inconsistent patch quality and error amplification at each stage.

\subsection{ContractTinker's Limitations}

\subsubsection{Vulnerability Example}

This section uses the high-risk vulnerability H-02 in the MarginSwap project as a case study, publicly audited by Code4Rena. Through a systematic analysis of this case, this paper reveals the primary issues associated with the ContractTinker method's use of a linear thinking chain~\cite{doi_10_1145_3691620_3695349}. This vulnerability is located in MarginRouter.sol, the core routing contract of the MarginSwap protocol, with the primary affected function being \texttt{crossSwapExactTokensForTokens}. This function serves as the protocol's core exchange entry point for users, performing the core business function of exchanging target tokens based on a fixed quantity of input tokens via AMM trading pair paths. Its core code is as follows:
\begin{lstlisting}[caption={MarginRouter main function}]
function crossSwapExactTokensForTokens(
    uint256 amountIn,
    uint256 amountOutMin,
    address[] calldata pairs,
    address[] calldata tokens
) external {
    uint256 fees = calculateFees(amountIn);
    uint256 amountInAfterFees = amountIn - fees;
    uint256[] memory amounts = UniswapStyleLib.getAmountsOut(
        amountInAfterFees, pairs, tokens
    );
    require(amounts[amounts.length - 1] >= amountOutMin, "Insufficient output");
    _swapExactT4T(amounts, pairs, tokens);
    registerTrade(amountIn, amounts[amounts.length - 1]);
}
\end{lstlisting}

An attacker can construct specific transaction parameters to exploit this vulnerability. This allows them to bypass the protocol's security checks and directly steal user assets from the project's Fund contract. Additionally, this attack method circumvents the withdrawal cooling-off period restrictions set by the protocol, enabling risk-free and instantaneous unauthorized fund transfers. According to the official repair recommendations from the Code4Rena platform and the results of the code review, the correct patch has two key points. These two points also serve as the core criteria for subsequently evaluating whether the patches generated by large language models are effective: Add a token uniqueness check at the beginning of the entry function \texttt{crossSwapExactTokensForTokens}. Move the withdrawal operation in the Fund contract to occur after the \texttt{startingBalance} is assigned and before interacting with the first trading pair~\cite{li2026psr2}.

\subsubsection{ContractTinker Method Flaws}

This paper conducts two experiments with ContractTinker using the same configuration parameters. The code and environment configurations are identical for both experiments. The differing output results from ContractTinker stem from the inherent generative randomness of large language models~\cite{olausson2023selfrepair, Hu2023LargeLM}. The experimental results show that the patch generated during the first run is completely ineffective. This patch fails to identify the root cause of the vulnerability and only makes code changes with no practical significance, thus failing to fix the H-02 vulnerability. The patch generated during the second run is conceptually correct but contains obvious logical issues. Specific problems include syntax errors, incomplete fix logic, and failure to address all root causes of the vulnerability. To address the poor quality and inconsistency of the vulnerability patches generated by ContractTinker, this paper conducts an end-to-end trace of the intermediate execution process within the large model's thought chain. ContractTinker breaks down the repair process into five steps in the thought chain: Q1-Extracting vulnerability-related elements, Q2-Root cause analysis of the vulnerability, Q3-Generating repair strategies, Q4-Code supplementation, and Q5-Generating code patches. As shown in Table~\ref{tab:q2_comparison}, we find that the large model's responses to step Q2 (root cause analysis of the vulnerability) in the thought chain differ fundamentally between the two runs.

\begin{figure}[htbp]
  \centering
  \includegraphics[width=1.0\linewidth]{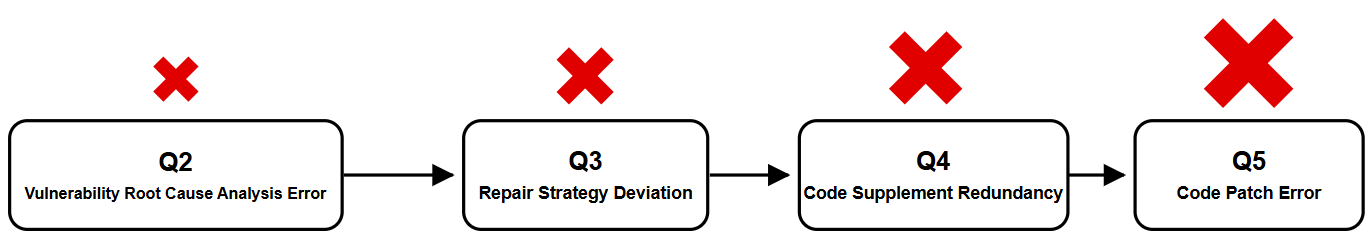}
  \caption{The error amplification effect of the linear thinking chain in ContractTinker.}
  \label{fig:fig2}
\end{figure}

\begin{table}[H]
  \caption{Comparison of Q2 output results from two runs.}
  \label{tab:q2_comparison}
  \centering
  \small
  \resizebox{\linewidth}{!}{%
  \begin{tabular}{p{3.7cm}p{4.5cm}p{4.5cm}}
    \toprule
    & \textbf{First Run of Q2} & \textbf{Second Run of Q2} \\
    \midrule
    Identification of root cause & Vague description; fails to identify the core issue & Hits Core Point 2: Balance Check Bypassed \\
    Description of Attack Logic & Error Description: Attack Entity & Accurate Reconstruction of Execution Sequence \\
    Attack Point Coverage & Focuses Only on Fake Reserves & Coverage of Fake Reserves and Execution Sequence \\
    \bottomrule
  \end{tabular}}
\end{table}
As shown in Figure~\ref{fig:fig2}, deviations in the root cause analysis of vulnerabilities during Phase Q2 will directly lead to the repair strategies generated in Phase Q3 being completely off track, ultimately resulting in low-quality patch code generated in Phase Q5 and repair outcomes that fall short of expectations. It can thus be concluded that the CoT method adopted by ContractTinker suffers from an inherent effect of error amplification at each stage, which is the core reason for the significant differences in the effectiveness of the patches generated in the two experiments and the overall poor quality of the patches. The characteristics of the linear chain of thought (single path, no backtracking, and no multi-solution verification) cause reasoning errors in earlier stages to amplify step by step along the chain, ultimately leading to output results that deviate completely from the repair objectives, and the stability of the output cannot be guaranteed.
\subsection{Theoretical Motivation}

Intuitively, ToT appears to be an ``upgraded version'' of CoT, so it is natural to consider using ToT instead of CoT to solve relatively complex problems. Previous studies have also used ToT to address vulnerability repair problems, but most have not explained the rationale for using ToT. Therefore, this section will elucidate the theoretical motivation for using ToT in this paper by constructing a simplified ToT model~\cite{fan2023automated, yang2022vulrepair, chen2022program}.

\subsubsection{Formal Theoretical Modeling}

\begin{figure}[H]
  \centering
  \includegraphics[width=0.8\linewidth]{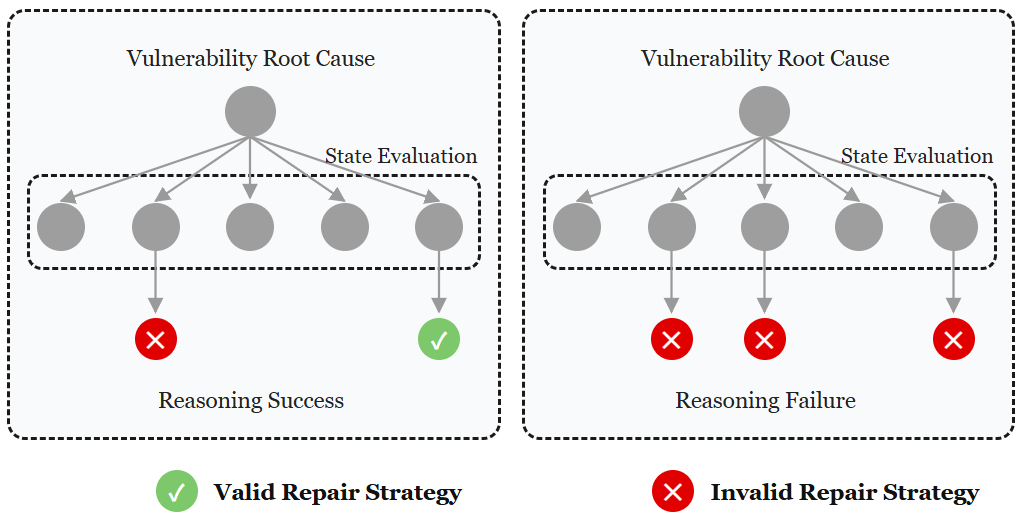}
  \caption{Criteria for Determining Successful Inference for CoT and ToT.}
  \label{fig:fig3}
\end{figure}

Take the intermediate step Q3-repair strategy generation in the ContractTinker method as an example. First, we establish a unified criterion for determining successful inference for both CoT and ToT: if the final output contains at least one valid repair strategy, the inference is deemed successful, as shown in Figure~\ref{fig:fig3}. For CoT, which generates only a single repair strategy, successful inference is equivalent to that generated strategy being a valid repair strategy~\cite{nye2022show, fu2023complexity, NEURIPS2023_271db992}.

Let the set of all valid repair strategies be denoted as the valid strategy space $S$. If a repair strategy $s \in S$, then that strategy is a valid strategy, otherwise, it is an invalid strategy. Let the probability that a repair strategy generated by the large language model in a single pass belongs to the valid space $S$ be denoted as $P_{s} = P(s \in S)$, where $1 > P_{s} > 0$. This probability is jointly determined by the large language model's code comprehension ability, domain knowledge coverage, and the quality of the contextual input.

For the baseline method ContractTinker, which employs single-path linear CoT reasoning, the repair strategy generation phase outputs only one candidate strategy. Therefore, the success probability of reasoning in this phase, $P_{CoT}$, is entirely determined by the effectiveness rate of that single strategy:
\begin{equation}
P_{CoT} = P_{s} \label{eq:cot}
\end{equation}

We consider using ToT multi-branch reasoning to address Q3-repair strategy generation. To simplify the theoretical model, ToT's thought generation and state evaluation are generated via independent and identically distributed sampling and evaluated independently, respectively. Meanwhile, the state evaluator is defined as a binary classifier responsible for determining the validity of the input repair strategy, outputting whether the strategy belongs to the valid space $S$. This paper uses two core metrics to evaluate the performance of the state evaluator:
\begin{align}
\eta_{tp} &= P(A \mid s \in S) \label{eq:eta_tp} \\
\eta_{fp} &= P(A \mid s \notin S) \label{eq:eta_fp}
\end{align}
where event $A$ indicates that the repair strategy $s$ is judged to be valid by the state evaluator, and $\eta_{tp}$ and $\eta_{fp}$, as defined in Eqs.~\eqref{eq:eta_tp} and \eqref{eq:eta_fp}, represent the true positive rate and the false positive rate, respectively.

Suppose that the idea generation phase generates a total of $k$ independent candidate repair strategies, and the state evaluator retains those it deems valid after screening. According to the criteria for a successful ToT inference, the conditions for inference failure are: among the $k$ candidate repair strategies, the number of valid repair strategies is 0, or the number of valid repair strategies is greater than 0, but all are erroneously rejected by the evaluator. Therefore, as shown in Eq.~\eqref{eq:pfail}, the probability of ToT inference failure is:
\begin{equation}
P_{fail} = \sum_{m=0}^{k}\binom{k}{m} \cdot p_{s}^{m} \cdot (1 - p_{s})^{k-m} \cdot (1 - \eta_{tp})^{m} = (1 - p_{s} \cdot \eta_{tp})^{k} \label{eq:pfail}
\end{equation}

Correspondingly, the probability of successful ToT inference, $P_{TOT}$, is derived from Eq.~\eqref{eq:ptot}:
\begin{equation}
P_{TOT} = 1 - P_{fail} = 1 - (1 - p_{s} \cdot \eta_{tp})^{k} \label{eq:ptot}
\end{equation}

Deriving the necessary and sufficient conditions for $P_{TOT} > P_{CoT}$ from Eqs.~\eqref{eq:cot} and \eqref{eq:ptot}, and simplifying the inequality yields $\frac{\ln(1 - p_{s})}{\ln(1 - p_{s} \cdot \eta_{tp})} < k$. For the sake of quantitative analysis, the number of sampled paths $k$ is set to a fixed value of 5. Setting $P_{s}$ and substituting the repair strategy success rate of 70.4\% reported by ContractTinker into the calculation yields $\eta_{tp}$. A success rate of just over 30.7\% is sufficient to satisfy $P_{TOT} > P_{CoT}$. This implies that even when using a state evaluator with extremely low performance, the ToT framework's inference success rate can still surpass that of single-path CoT inference.

\subsubsection{Analysis of Results}

\begin{figure}[H]
  \centering
  \includegraphics[width=0.85\linewidth]{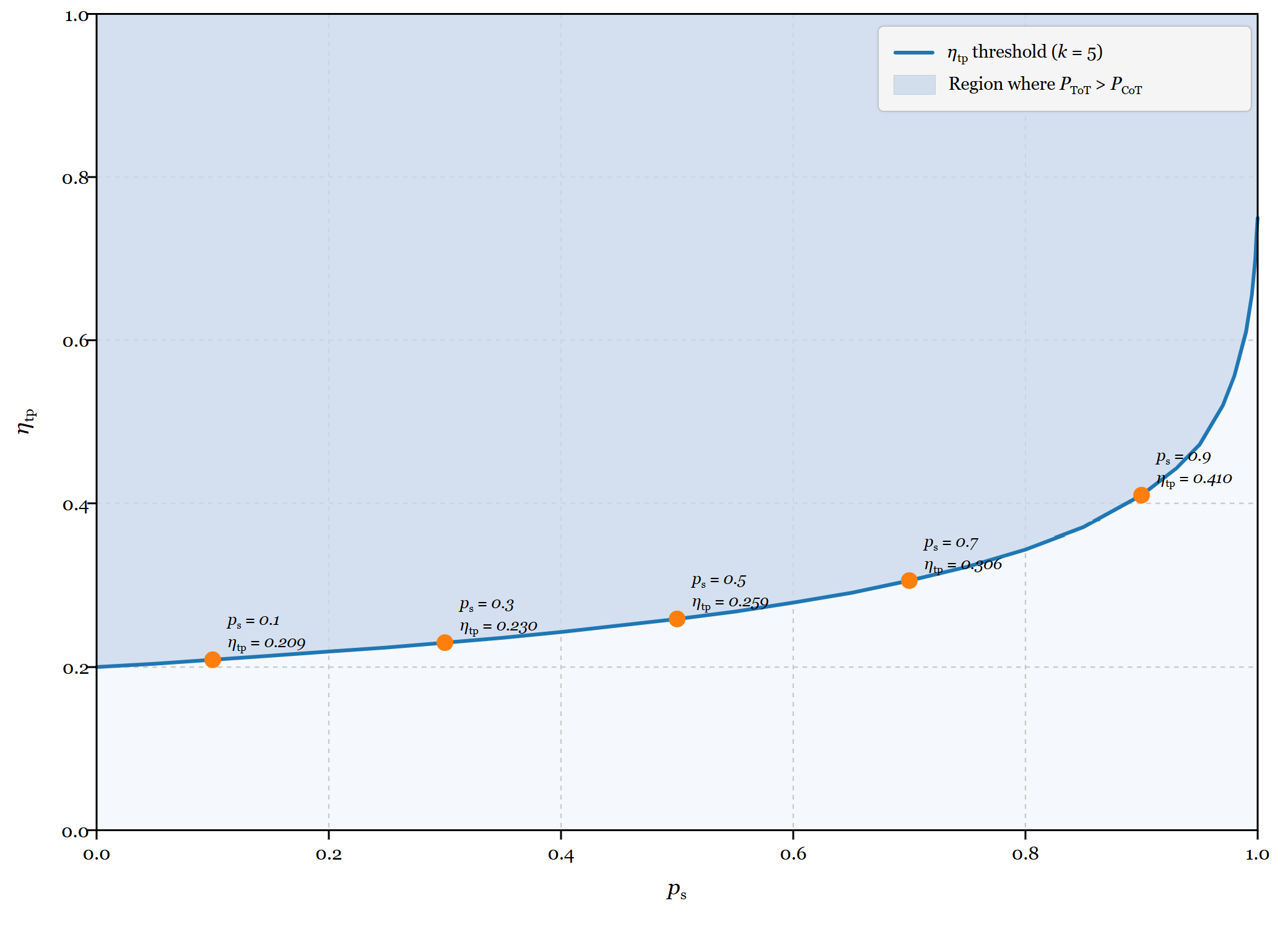}
  \caption{Relationship between the true positive rate of the state evaluator and the success rate of the single-repair strategy when $k=5$.}
  \label{fig:fig4}
\end{figure}

\begin{figure}[H]
  \centering
  \includegraphics[width=0.85\linewidth]{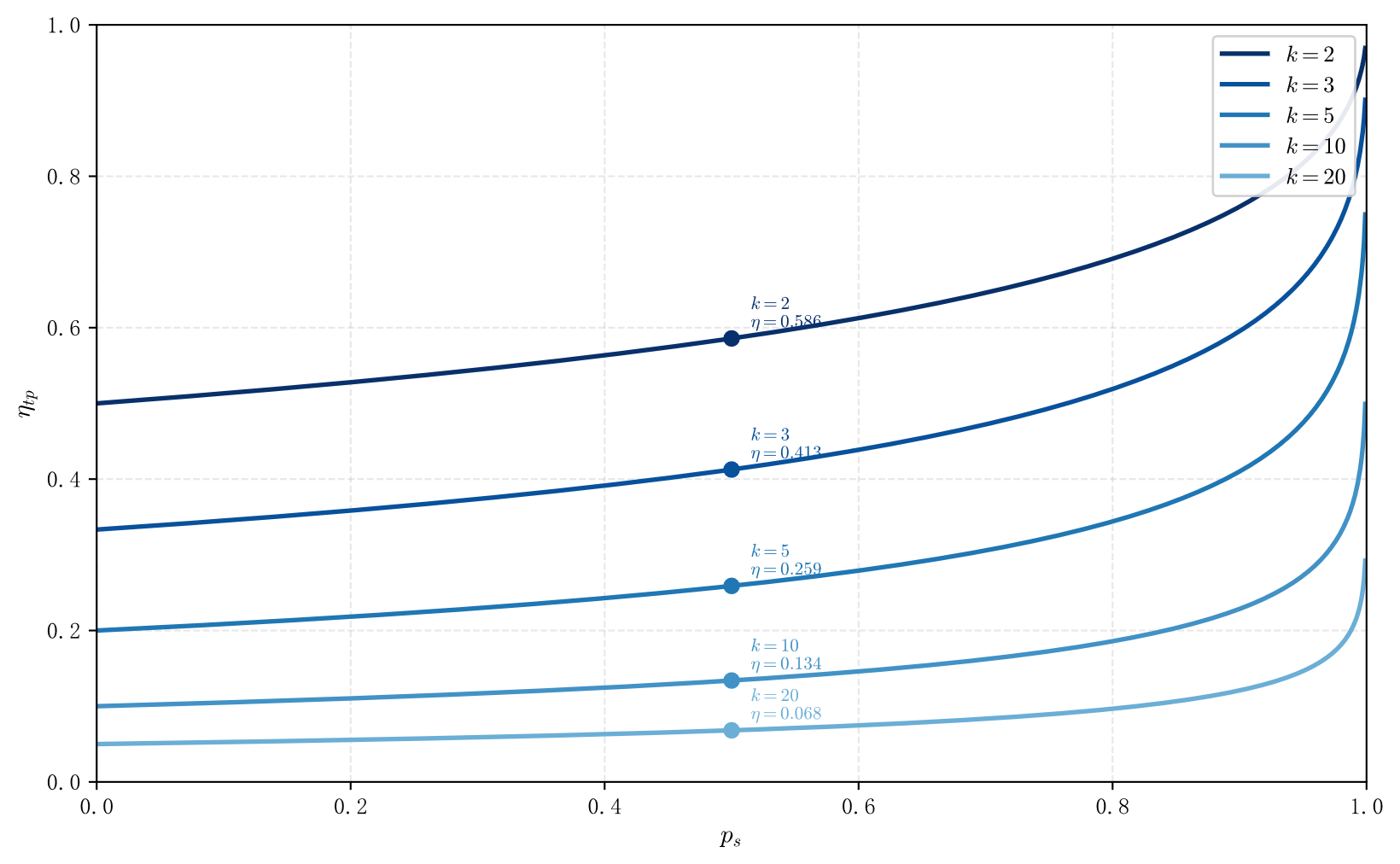}
  \caption{Relationship between the true positive rate of the state evaluator and the success rate of the single-repair strategy for different values of $k$.}
  \label{fig:fig5}
\end{figure}

Figure~\ref{fig:fig4} illustrates the relationship between $\eta_{tp}$ and $P_{s}$ for $k=5$. The blue region in the figure represents the critical value of $\eta_{tp}$ when $P_{TOT} > P_{CoT}$ holds, while the light blue region indicates the area where $P_{TOT} > P_{CoT}$. Figure~\ref{fig:fig5} illustrates the relationship between $\eta_{tp}$ and $P_{s}$ for different values of $k$~\cite{suzgun2023challenging}. Three valuable conclusions can be drawn from the figure: (1) The poorer the performance of single-path CoT, the easier it is for ToT to outperform CoT. When the success rate of a single repair strategy generated by CoT is 0.3, ToT's state evaluator only needs a true positive rate of 23\% to outperform CoT. (2) ToT is suitable for the vast majority of scenarios. Figure~\ref{fig:fig4} shows that the growth trend of the curve exhibits extremely strong nonlinear characteristics. As $P_{s}$ increases from 0.1 to 0.7, the critical value $\eta_{tp}$ rises only slowly from 20.9\% to 30.6\%, representing a very gradual increase. In real-world vulnerability repair scenarios, it is virtually impossible for a single CoT-generated repair strategy to achieve a success rate exceeding 90\%. This means that the vast majority of vulnerability scenarios fall within ToT's low-threshold comfort zone, where an evaluator with even basic discriminative capabilities can consistently achieve performance improvements. (3) The larger the number of sampled paths $k$, the less stringent the performance requirements for the evaluator under ToT. However, in practical applications, increasing $k$ leads to a simultaneous rise in inference costs and processing pressure on subsequent nodes, so a balance must be struck between performance gains and engineering overhead. The above conclusions indicate that, in the vast majority of scenarios, replacing CoT with ToT is effective. The theoretical analysis above is based on a simplified ToT framework, if tailored and adapted for specific tasks, ToT's performance can be further improved.

\subsection{ToT for Smart Contract Vulnerability Repair}

The core framework of ToT consists of four key modules: Thought Decomposition, Thought Generator, State Evaluator, and Search Algorithm. This section will detail the design and construction of these four modules one by one, and elucidate the design logic and rationale for each module's adaptation to the smart contract vulnerability repair scenario~\cite{wang2023plan, yao2023react, hao2023reasoning, NEURIPS2023_271db992, koh2024tree}. Thought Decomposition. ToT leverages the characteristics of the problem to design and decompose intermediate thought steps. An appropriate thought unit enables the large language model not only to generate promising and diverse candidate approaches but also to evaluate the potential of those approaches for solving the problem. This paper adopts the decomposition of vulnerability repair tasks from ContractTinker, which divides the task into four parts: extracting vulnerability-related information, decomposing the attack process, proposing repair strategies, and generating code patches. Thought Generator. Thought generators can be categorized into two types based on their generation strategies: proposed prompt generation and independent and identically distributed sampling generation. Proposed prompt generation is suitable for task scenarios with a limited thought space, while sampling generation is better suited for scenarios with an open and rich thought space. This study adopts the proposed prompt generation strategy. The reason for this is that the core repair points and feasible solutions for smart contract vulnerability repair all lie within a finite search space, if the independent identically distributed sampling generation strategy were used, it would generate a large amount of redundant and repetitive text content, thereby significantly reducing the efficiency of reasoning and search. State Evaluator. Given a set $S$ of leading states, the state evaluator assesses the progress these states make toward solving the problem, serving as heuristic rules (used by the search algorithm to determine which states to continue exploring and in what order). Heuristic rules are a standard method for solving search problems; this paper uses a language model to carefully reason about states. Similarly, the state evaluator can be divided into two strategies: independent evaluation and cross-state voting evaluation. This paper uses the latter as the evaluation strategy because, in vulnerability repair tasks, the text generated by the idea generator is difficult to quantify directly. Search Algorithms. In ToT, different search algorithms can be flexibly inserted or removed based on the tree structure. Two relatively simple search algorithms are breadth-first search (BFS) and depth-first search (DFS). BFS maintains the $b$ most promising states at each step. This algorithm is suitable for tasks with limited tree depth, such as the 24-point game and creative writing, where the initial thought steps can be evaluated and pruned down to a small set. DFS prioritizes exploring the most promising states until a final output is generated, or until the state evaluator determines that the problem cannot be solved from the current state $s$. In the latter case, the subtree originating from $s$ is pruned, and search resources are backtracked to $s$'s parent state to continue exploration. This paper uses BFS as the search algorithm for ToT because the ToT used here has a limited depth (only three levels), and it is difficult for the state evaluator to determine whether a state $s$ generated by DFS can solve the problem.
\section{Method}

\subsection{Method Overview}

\begin{figure}[htbp]
  \centering
  \includegraphics[width=1.0\linewidth]{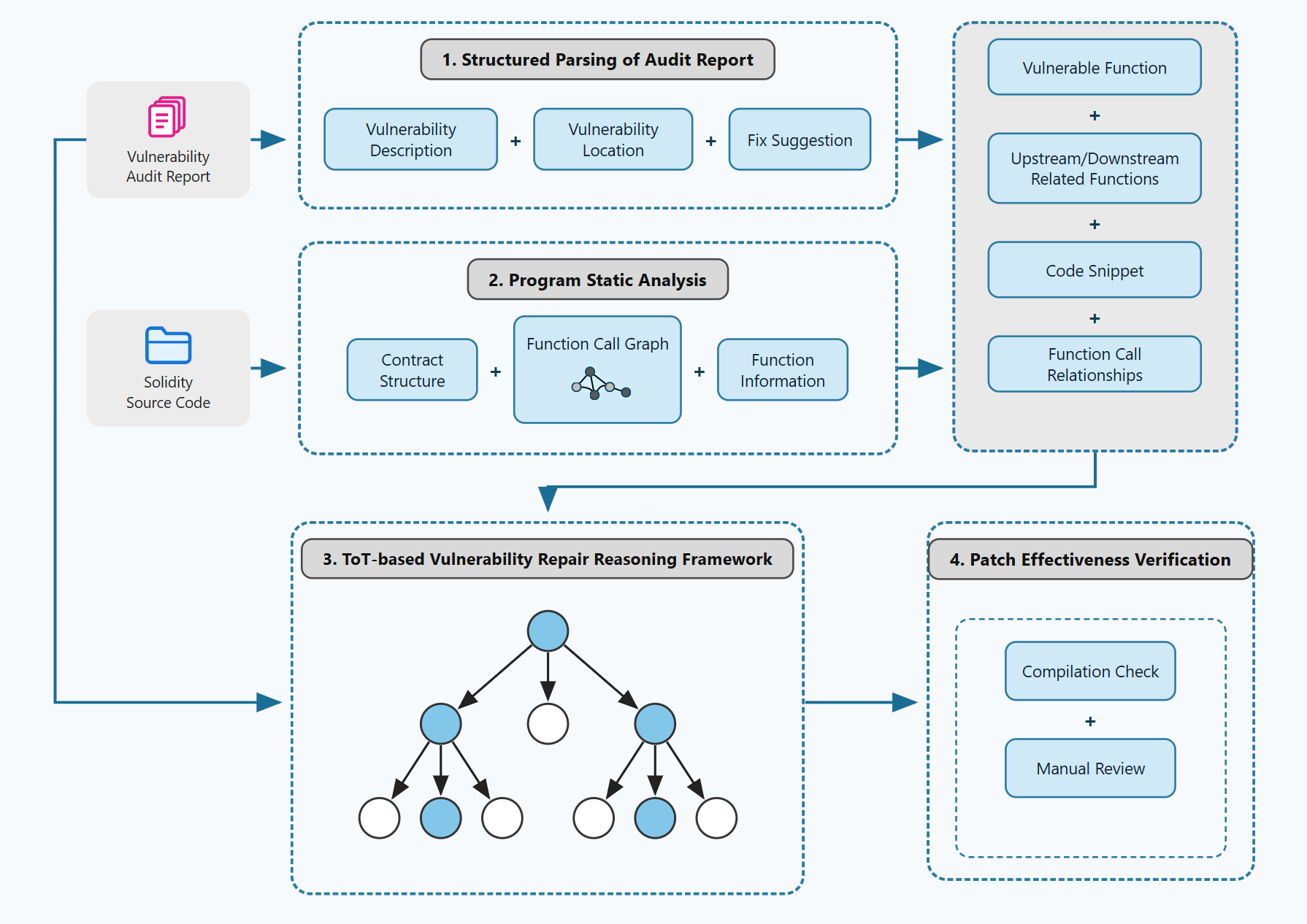}
  \caption{The overall framework of the proposed smart contract vulnerability repair method based on ToT.}
  \label{fig:fig6}
\end{figure}

One of the core challenges in the automatic repair of smart contract vulnerabilities lies in the fact that audit reports are mostly unstructured natural language text, making it difficult to directly correlate them with the syntax and semantics of the contract code. However, large language models relying on a single thought chain tend to get stuck in local optima, leading to one-sided repair strategies. To address these issues, this paper proposes a method for the automatic repair of smart contract vulnerabilities that integrates structured document parsing, static code analysis, and multi-branch ToT reasoning. The overall framework is shown in Figure~\ref{fig:fig6}.

This framework consists of four main steps: The first step is to parse the audit report. Key information (such as vulnerability descriptions, locations, and repair recommendations) is extracted from the natural language audit report. This addresses the issue of inconsistent report formats. The second step is static analysis of the program. The target smart contract project is comprehensively parsed to extract code context information, such as the structure of the contract and functions, as well as the call relationships between functions. This information provides accurate semantic support for subsequent fixes. The third step involves vulnerability reasoning and repair based on ToT. Using the previously extracted structured vulnerability information and program context as input, the ToT framework guides a large language model to perform multi-angle root-cause analysis of the vulnerability, generating repair strategies and code patches. The fourth step is to verify the effectiveness of the patch. The generated patch code is integrated into the original contract project and fully compiled using the corresponding version of the Solc compiler and dependency environment to check for basic issues such as syntax errors, missing dependencies, and undefined variables. Finally, personnel with experience in smart contract security auditing manually verify the patch's security, effectiveness, and compatibility to ensure it can be used safely in real-world scenarios. This framework balances the flexibility of natural language understanding with the precision of program analysis. By leveraging ToT multi-branch reasoning to overcome the limitations of a single thought chain, it achieves end-to-end automated repair of vulnerable code.

\subsection{Structured Analysis of Audit Reports}

Audit reports serve as a crucial basis for vulnerability repair. However, audit reports primarily use natural language, and their content is unstructured. This unstructured nature makes it difficult for machines to directly understand the report's content. The goal of this module is to convert unstructured audit reports into structured data that machines can process. This conversion process primarily involves two tasks: extracting vulnerability location information and extracting semantic information about the vulnerabilities.

\subsubsection{Vulnerability Location Extraction}

Different audit firms use varying formats to describe vulnerability locations. To address this issue, we have designed a multimodal regular expression matching strategy. This strategy can automatically convert unstructured vulnerability location text into structured data. Vulnerability locations in smart contracts are typically represented by the .sol filename and line number. The strategy described in this paper first defines a set of regular expressions that cover all the formats mentioned above. This allows for the accurate extraction of filenames and corresponding line numbers from the text. After extraction, we deduplicate and standardize the raw match results. All results are uniformly output as ``filename-line number'' key-value pairs. If only a filename is matched and no line number is found, the line number field is marked as `None'. Finally, this paper constructs a vulnerability location dictionary. This dictionary uses filenames as keys and the corresponding lists of line numbers as values. This completes the structured storage of vulnerability location information, providing a unified data interface for subsequent vulnerability localization and code fixes.

\subsubsection{Extraction of Vulnerability Semantic Information}

The semantic information of smart contract vulnerabilities primarily consists of two parts: vulnerability descriptions and repair recommendations. Vulnerability descriptions explain how the vulnerability arises, which areas it affects, and technical details such as potential attack vectors. Repair recommendations document the corrective measures proposed by auditors. Report formats vary significantly across different audit organizations, and vulnerability descriptions and repair recommendations are often intermixed. To address this issue, we design a set of text segmentation and cleaning strategies capable of converting this unstructured text into structured data. This strategy consists of several steps. First, text cleaning is performed to automatically remove content unrelated to the vulnerability from the reports, such as redundant comments, GitHub links, and version information. This reduces noise and improves the accuracy of subsequent processing. Next, a keyword-based semantic segmentation method is used to accurately identify the boundaries between the ``vulnerability description'' and ``repair recommendations'' sections. If no clear segmentation keywords are found, the entire text segment is classified as a vulnerability description to prevent the omission of important information. The segmented text is then filtered to remove non-essential information, such as submitter details, audit dates, and report numbers, retaining only core information related to the vulnerability's technical details and repair solutions. After these processing steps, a structured dictionary is generated containing three fields: vulnerability description, vulnerability location, and repair recommendation. This dictionary provides a unified natural language input for subsequent large language model inference. It is important to note that the repair recommendations are used only in the final evaluation phase to verify whether the code patches generated by the model are effective; they do not participate in the model's inference process.

\subsection{Static Program Analysis}

The purpose of static code analysis is to identify the structure and meaning of the contract code. This information helps large language models fix the code more accurately. This module uses the Slither tool to perform program static analysis. The entire process consists of three main steps. The first step is to parse all contract files to understand their structure. The second step is to generate a function call graph that illustrates the call relationships between functions~\cite{li2024interaction}. The third step is to identify relevant functions and collect the necessary code snippets~\cite{tsankov2018securify, mossberg2019manticore, li2025penetrating, zhang2022cross}.

\subsubsection{Comprehensive Parsing of Contract Files and Structures}

First, the target project directory is scanned to identify all contract files. Then, the Slither tool is used to parse these contract files one by one. During the parsing process, the program establishes a mapping relationship that links contract names to their corresponding contract objects. Each contract object contains important information, such as which functions the contract has, the source code for each function, and which modifiers the functions use (e.g., public, external, view, etc.). In this way, we organize the structural information of contracts and functions into structured data for convenient use later.

\subsubsection{Generation of Function Call Graphs}

A call graph records the call relationships between various functions within a smart contract. It clearly displays the control flow and data transfer paths within the contract. This paper uses Slither, a static analysis tool, to generate standard DOT-format call graph files for each .sol source file. After obtaining the raw call graph data, the DOT files are first parsed and preprocessed. This step primarily involves removing redundant information, such as numerical suffixes in node names and anonymous function identifiers, and converting all function names to the standard form defined in the source code. Next, the standardized function names are used as graph nodes, and the call relationships between functions are used as edges to construct the contract's function interaction graph. This graph is saved in a dictionary structure in the format ``source function-list of target functions,'' which facilitates subsequent queries of the call context for any given function. Finally, duplicate edges in the graph are globally deduplicated to ensure that each function call relationship is recorded only once, thereby preventing duplicate data from affecting subsequent analysis.

\subsubsection{Function Matching and Code Snippet Collection}

To accurately map vulnerability descriptions to actual code, this paper designs a multi-granularity matching method. This method matches contract and function keywords extracted from vulnerability reports with the contract structure and function call graph obtained through static analysis. Since report formats vary significantly across different audit organizations and vulnerability descriptions contain a large amount of natural language, traditional keyword matching struggles to accurately locate relevant code. Therefore, we use a large language model to extract contract and function keywords, leveraging its semantic understanding capabilities to improve identification accuracy~\cite{li2026ckgllm}. The matching process consists of three steps. The first step is contract-level filtering, which identifies core contracts containing keywords to narrow the scope of analysis. The second step involves matching corresponding core functions within the core contracts, then using the function call graph to locate associated functions directly called by the core functions, thereby covering all code potentially affected by the vulnerability. The third step is to extract the complete source code of the core and associated functions to serve as the code context for the large language model when fixing the vulnerability. This multi-granularity matching mechanism enables automatic mapping from natural language vulnerability descriptions to precise code context. It addresses two issues: first, it avoids feeding the entire contract code to the large language model, which would result in an overly large context window; second, it minimizes the interference of irrelevant information on the effectiveness of the fix. At the same time, it ensures that the code context required for the fix is complete.

\subsection{ToT-Based Vulnerability Fix Reasoning Framework}

\begin{figure}[htbp]
  \centering
  \includegraphics[width=0.72\linewidth]{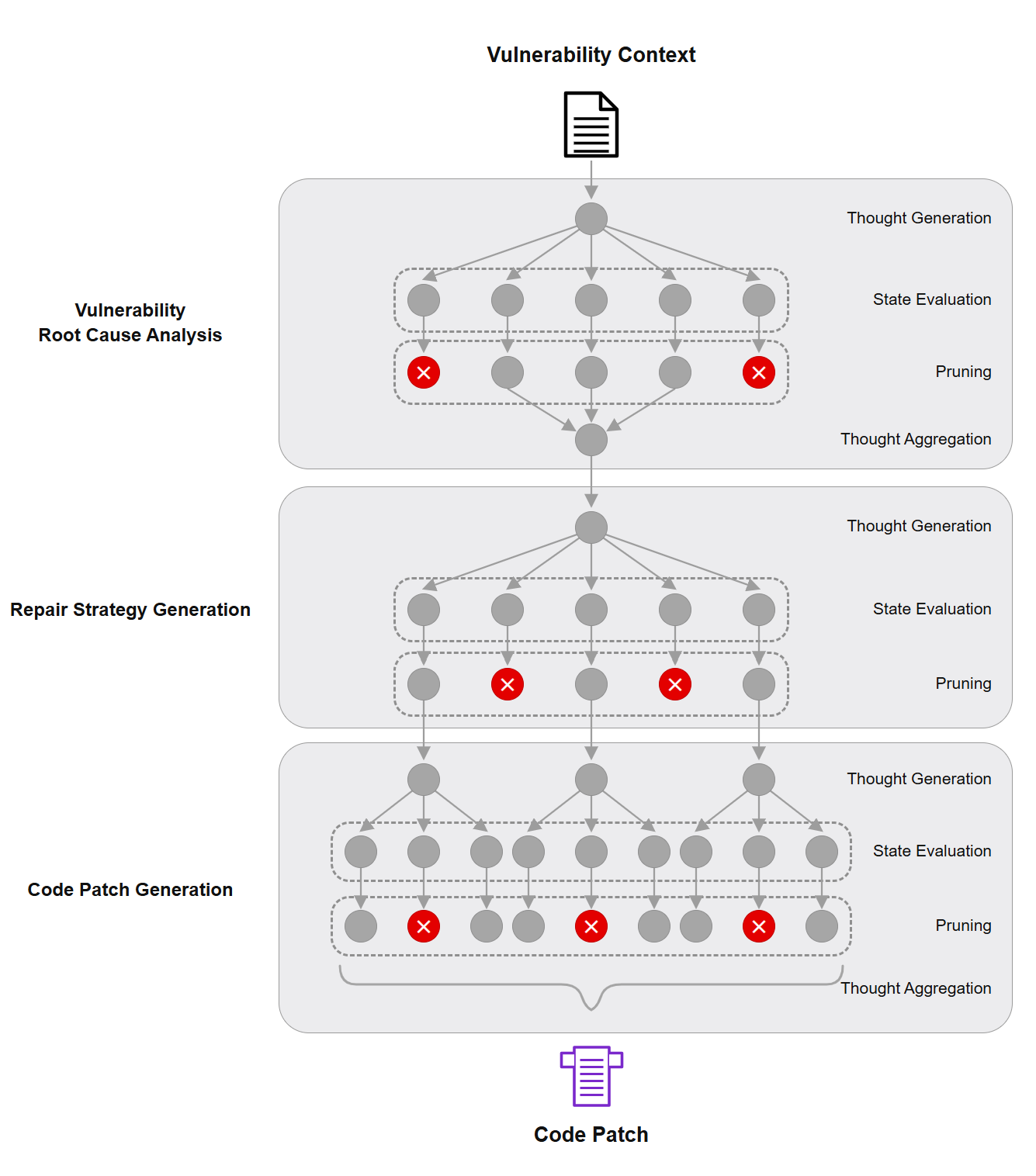}
  \caption{ToT Reasoning Repair Process.}
  \label{fig:fig7}
\end{figure}

As mentioned above, this paper combines the ToT framework with the vulnerability repair task to design a three-stage vulnerability repair reasoning process, as shown in Figure~\ref{fig:fig7}. The overall repair process is divided into three parts: root cause analysis of the vulnerability, repair strategy generation, and code patch generation. Among these, root cause analysis employs a self-consistent thought chain, while repair strategy generation and code patch generation utilize a multi-branch thought tree~\cite{Huang2023AgentCoderMC, zhou2023least, zhang2023automatic, madaan2023selfrefine, long2023large, CHOI2025785}. During the vulnerability root cause analysis phase, CoT-SC is used to select the most consistent answer from multiple inference branches. If an error occurs in an early inference step, it will be amplified throughout the entire process, rendering all subsequent steps ineffective. Therefore, when analyzing the root cause of a vulnerability, the most important consideration is ensuring that the conclusions are reliable and consistent, rather than exploring multiple possible inference paths. Specifically, during the root cause analysis phase, the large language model first generates five distinct analytical approaches. These approaches are independent of one another, and each has its own characteristics. Next, the state evaluator scores each approach. We select the three highest-scoring approaches. Finally, the large language model fuses and optimizes these three high-scoring approaches to produce a unified reconstruction of the attack flow and a conclusion regarding the root cause of the vulnerability. During the repair strategy generation phase, the ToT multi-branch reasoning framework is used to explore multiple viable repair solutions. First, the large language model generates five different candidate repair strategies. After the state evaluator scores these strategies, the top three high-quality strategies are selected. During the code patch generation phase, three different candidate patches are generated for each selected repair strategy. The state evaluator scores each of the three patches corresponding to every strategy and selects the two highest-scoring patches for each strategy, resulting in a total of six candidate patches. Finally, the large language model fuses and optimizes these six patches to generate a final code patch solution that is logically sound and compliant with business specifications.

\subsection{Patch Validity Verification}

This phase serves as the final verification step in the vulnerability repair process. Its primary objectives are to confirm whether the patch compiles correctly, whether it actually fixes the vulnerability, and whether it affects business functionality. The entire verification process consists of two steps: first, a compilation check, followed by manual verification. During the compilation check phase, the specific Solidity compiler version specified in the contract source code is first identified. Next, the patch code generated by the ToT framework is integrated into the original contract project. The entire contract is then compiled using the Slither tool. This step primarily checks for basic issues such as syntax errors, undefined variables, missing dependencies, mismatched function signatures, and compiler version incompatibilities. During the manual verification phase, personnel with experience in smart contract security auditing conduct a final review of the patches that passed the compilation check, following systematic security assessment practices~\cite{zhang2025penetration}. This review references the structured output from `patch\_generate', which includes information such as the root cause analysis of the vulnerability, attack paths, core function code, and code comparisons before and after the patch~\cite{olausson2023selfrepair, shinn2023reflexion}. The review focuses on three key aspects: First, whether the patch fully addresses the root cause of the vulnerability and blocks all attack paths identified by the ToT framework. Second, whether the patch adheres to the principle of minimal modification and does not compromise the contract's original business logic and functionality. And third, whether the patch introduces new security risks or logical defects. Finally, based on the review results, patches are classified into four levels: ``Effective and Implementable,'' ``Requires Minor Modifications,'' ``Requires Logical Optimization,'' and ``Ineffective Fix.'' This provides users with clear implementation guidance.
\section{Evaluation}

\subsection{Experimental Setup}

This paper selects 50 high-risk vulnerabilities from the Code4Rena platform as the experimental dataset. Each vulnerability is accompanied by an officially published vulnerability description and repair recommendations. Regarding evaluation metrics, this paper primarily employs two types of evaluation systems to comprehensively assess model performance. The first is the fix strategy success rate, calculated by comparing the fix strategies generated by the model with the official recommendations. The fix strategy success rate refers to the proportion of samples where the model's generated fix strategies fully cover the official recommendations. The Top-3 fix strategy success rate refers to the proportion of samples where at least one of the top three fix strategies generated by the model matches the official recommendation. The second category is the quality assessment of code patches. Since the official sources do not provide specific fix code, we employ three methods to verify the accuracy of the patches: automated checking by Qwen3-Max-Thinking, manual review, and compilation verification. During the compilation verification phase, the generated code patches are integrated into the original smart contract to verify whether the contract could compile successfully. During the Qwen3-Max-Thinking automated inspection and manual review phases, the functional correctness of the generated patches is verified using the Qwen3-Max-Thinking large language model and manual review, respectively. Based on the above validation strategy, all generated patches are classified into the following four categories: fully valid patches, functionally valid patches requiring minor adjustments, patches with defects in core logic requiring refactoring, and invalid patches that cannot fix the vulnerability~\cite{jimenez2024swebench, chaliasos2024smart, bose2022smartbugs, tufano2022empirical}. The experiment uses GPT-4o-mini as the base large language model, responsible for both the reasoning generation and state evaluation tasks. All inference and evaluation are performed via calls to OpenAI's official API. In the experiment, the model's temperature parameter is set to 1.0. The experiment ran on an Alibaba Cloud Elastic Cloud Server. The server configuration consists of a 2-core CPU and 2 GiB of memory, running the 64-bit Ubuntu 22.04 operating system. Smart contracts are compiled using the solc compiler, which automatically selects the appropriate Solidity version based on the one specified in the contract source code. Static analysis and compilation checks for the contracts are performed using the Slither tool.

\subsection{Results}

\begin{figure}[htbp]
  \centering
  \includegraphics[width=0.64\linewidth]{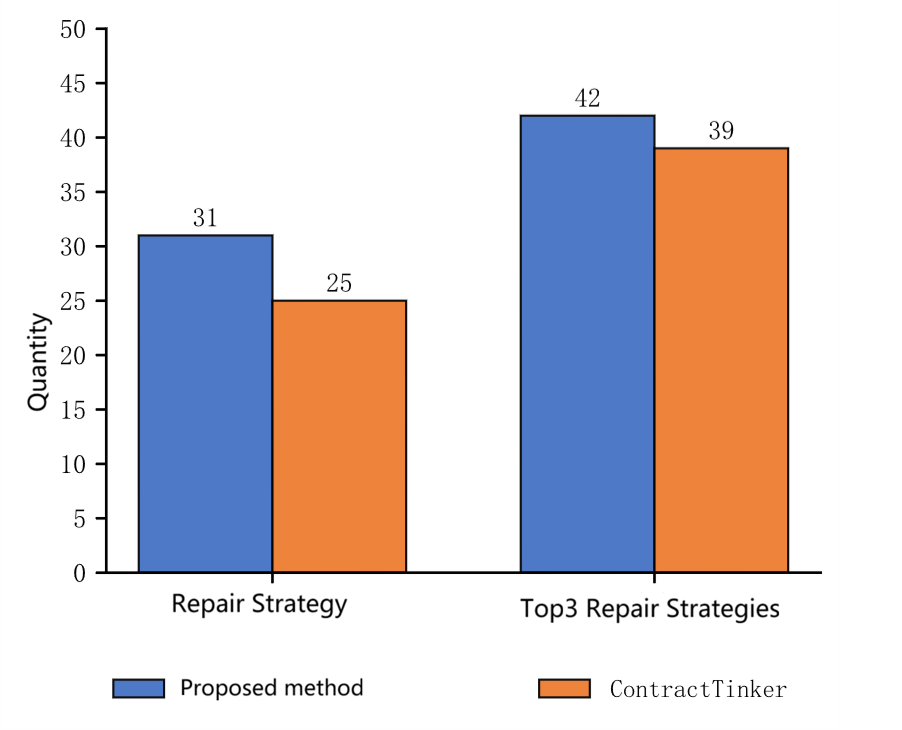}
  \caption{Comparison of the success rates of repair strategies using different methods.}
  \label{fig:fig8}
\end{figure}

\begin{figure}[H]
  \centering
  \includegraphics[width=0.64\linewidth]{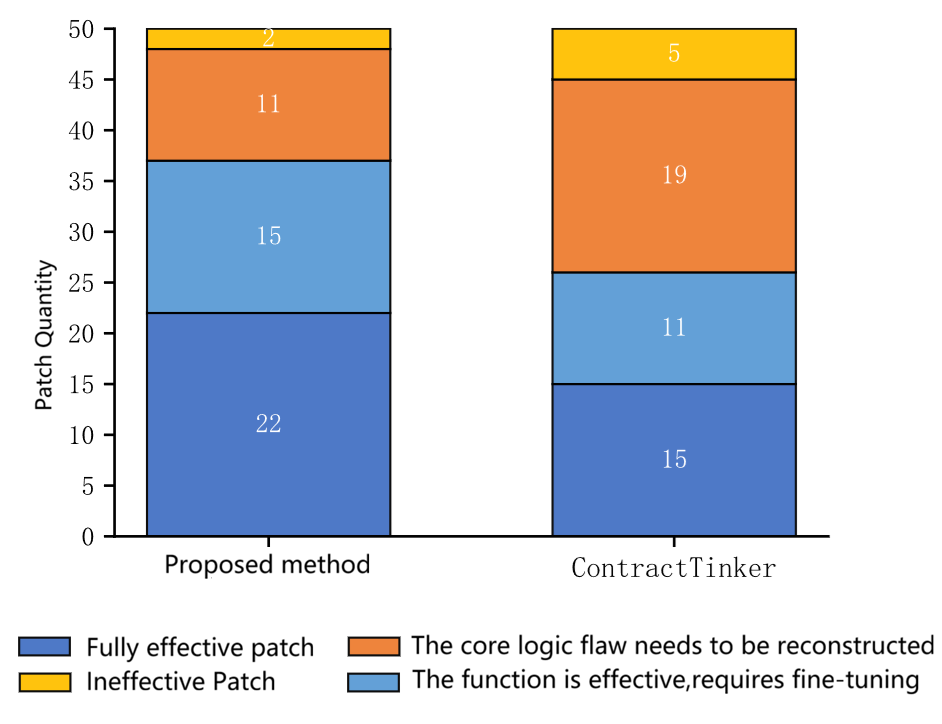}
  \caption{Comparison of the quality distribution of patches generated by different methods.}
  \label{fig:fig9}
\end{figure}

This paper conducts comparative experiments on 50 real-world high-risk vulnerability datasets selected from the Code4Rena platform. Using ContractTinker as the baseline method, the effectiveness of the proposed thought-tree-based smart contract vulnerability repair method is verified in terms of both repair strategy success rate and patch quality. The experimental results are shown in Figure~\ref{fig:fig8} and Figure~\ref{fig:fig9}. The experimental results indicate that the method proposed in this paper significantly outperforms the baseline method on both key metrics: the accuracy of repair strategy generation and the quality of the generated patches. In terms of repair strategy success rate, the method proposed in this paper achieves a success rate of 62\%, meaning that 31 out of 50 vulnerabilities can be correctly repaired. This result is 12 percentage points higher than ContractTinker's 50\%. The Top-3 repair strategy success rate reaches 84\%, meaning that for 42 out of 50 vulnerabilities, at least one of the top three strategies is correct. This result represents a 6 percentage point improvement over the baseline method's 78\%. These data demonstrate that the multi-branch reasoning framework based on a ``thought tree'' used in this paper is indeed effective, as it avoids the problem of errors being amplified step by step in a linear chain of reasoning. Through multi-path exploration and state-based pruning, our method enhances the reliability of vulnerability root cause analysis and increases the coverage of repair strategies. In terms of patch quality, the method described in this paper generates 22 fully valid patches, accounting for 44\% of the total. This figure is significantly higher than ContractTinker's 15 patches, or 30\%. The method also generates 15 patches that are functionally valid but require minor adjustments, accounting for 30\%. This also outperforms the baseline method's 11 patches, or 22\%. In contrast, the method described in this paper generates only 11 patches with flawed core logic requiring refactoring and only 2 invalid patches. Both of these figures are far lower than ContractTinker's 19 and 5, respectively. The corresponding proportions also decrease from the baseline method's 38\% and 10\% to 22\% and 4\%. These data indicate that our method not only increases the proportion of high-quality patches that can be used directly but also significantly reduces the number of low-quality and invalid patches. The overall quality and practicality of the patches have improved significantly. This is because our method provides the large model with accurate code context through a static code analysis module. At the same time, the multi-solution exploration and evaluation mechanism of the thought-tree framework effectively reduces logical errors caused by hallucinations generated by the large model. The patches generated in this way can not only accurately block attack paths but also preserve the contract's original business logic to the greatest extent possible~\cite{jiang2023impact}.

\subsection{Case Study}

In this section, we select the H-01 vulnerability from the MarginSwap project and the H-02 vulnerability described in Section~3 as research subjects. By comparing the differences between the ContractTinker tool and the method proposed in this paper in terms of code patch generation and repair strategies, we validate the effectiveness of our method. The selection of these two vulnerabilities as test cases is primarily based on the following considerations: First, the H-01 and H-02 vulnerabilities are highly representative and significant within the MarginSwap project. Second, their root causes are highly coupled with the project's core business logic, resulting in a high degree of complexity. And finally, evaluating the effectiveness of fixes for such vulnerabilities, which are deeply intertwined with business logic, can more comprehensively and objectively reflect the adaptability and superiority of the method proposed in this paper in real-world application scenarios~\cite{zhou2023sok, li2026defi}.

\subsubsection{H-01 Vulnerability}

The H-01 vulnerability is a composite security vulnerability that combines a reentrancy attack with price oracle manipulation~\cite{qin2021empirical, liu2021towards}. By deploying a malicious contract, an attacker cleverly manipulates price data during transaction execution and exploits the reentrancy mechanism to trick the system into recording false, high transaction profits, ultimately leading to the systematic draining of the protocol's funds. The core code for the H-01 vulnerability is as follows:

\begin{lstlisting}[caption={Vulnerable function in MarginRouter.sol}]
function crossSwapExactTokensForTokens(
    uint256 amountIn,
    uint256 amountOutMin,
    address[] calldata pairs, // controllable by attacker
    address[] calldata tokens,
    uint256 deadline
) external ensure(deadline) returns (uint256[] memory amounts) {
    // 1. Price retrieval: external contract (manipulable)
    amounts = UniswapStyleLib.getAmountsOut(amountIn - fees, pairs, tokens);
    // 2. Risk verification: based on manipulated price
    registerTrade(msg.sender, tokens[0], tokens[tokens.length - 1],
                  amountIn, amounts[amounts.length - 1]);
    // 3. Execution: attacker can reenter here
    _swapExactT4T(amounts, amountOutMin, pairs, tokens);
}
\end{lstlisting}

The fix for this vulnerability must address two key issues simultaneously: price data validation and reentrancy protection. The official vulnerability audit report from Code4Rena provides the following repair recommendations: Add reentrancy protection: Apply OpenZeppelin's ReentrancyGuard to all external functions of MarginRouter. Restructure the trading process: Remove the estimation step; execute the actual trade first, then call \texttt{registerTrade} with the actual trade amount. Table~\ref{tab:h01_strategies} shows the top three repair strategies proposed by ContractTinker and this paper. The results show that only one of ContractTinker's reentrancy protection strategies meets the official requirements. The other two strategies have obvious issues. The first strategy, which adjusts the execution order, violates the official ``trade first, then register'' repair logic. The second strategy, which adjusts the balance check, merely replicates the logic already present in the original code and fails to address the core vulnerability of duplicate authorization. Overall, ContractTinker's repair solution has significant limitations. In contrast, the three-pronged repair strategy proposed in this paper has passed validity verification: it not only incorporates a reentrancy protection mechanism consistent with official recommendations but also addresses the attack at its source by adding two critical defensive measures, address whitelist verification and oracle reserve value validation. The address whitelist blocks the attack path of forged trading pairs at the input layer, while oracle verification resolves the fundamental issue of estimation values relying on untrusted contracts at the data layer. Together, these three components form a comprehensive, multi-layered defense system. Compared to ContractTinker's partially effective and partially flawed fix, the method proposed in this paper covers a broader attack surface, provides a more thorough fix, and significantly enhances security and robustness.
\begin{table}[H]
  \caption{Comparison of the Effectiveness of Different Repair Strategies for the H-01.}
  \label{tab:h01_strategies}
  \centering
  \small
  \resizebox{\linewidth}{!}{%
  \begin{tabular}{p{2.5cm}p{2.5cm}p{3cm}>{\centering\arraybackslash}p{1.2cm}p{3.5cm}}
    \toprule
    \multicolumn{1}{c}{\textbf{Method}} & \multicolumn{1}{c}{\textbf{Strategy}} & \multicolumn{1}{c}{\textbf{Description}} & \multicolumn{1}{c}{\textbf{Valid}} & \multicolumn{1}{c}{\textbf{Analysis}} \\
    \midrule
    \multirow{3}{*}{ContractTinker} & Reentrancy Protection & Add a lock to prevent reentrancy & $\surd$ & Complies with official recommendations \\
    & Adjust Execution Order & Validate before transaction & $\times$ & Contrary to official recommendations \\
    & Adjust balance check & Verify balance after transaction & $\times$ & Does not address core vulnerability \\
    \midrule
    \multirow{3}{*}{Our Method} & Reentrancy Protection & Add a lock to prevent reentrancy & $\surd$ & Complies with official recommendations \\
    & Introduce an oracle & Retrieve actual reserve values & $\surd$ & Official alternative \\
    & Address whitelist & Allow only trusted transactions & $\surd$ & Official alternative \\
    \bottomrule
  \end{tabular}}
\end{table}

Table~\ref{tab:h01_ct_patches} and Table~\ref{tab:h01_our_patches} compare the results of validity verification for code patches from the two types of repair solutions. The results show that of the two patches generated by ContractTinker, only one meets official requirements. The other patch not only fails to address the core vulnerability, but also causes a functional issue of transaction failure due to incorrect exchange of the fund transfer order. It is evident that ContractTinker's overall repair solution has serious shortcomings. In contrast, all three patches proposed by this paper offer practical protective value: the reentrancy guard patch fully complies with official core recommendations and can directly block the path for repeated authorization attacks. The prophecy machine amount calculation patch is an effective alternative to official recommendations, addressing at the data layer the root cause of reliance on untrusted contracts for estimated values. The reserve verification patch, supported by trusted data sources, further defends against reserve forgery attacks. Compared to ContractTinker's partially effective solutions, which involve disruptive modifications, the method presented in this paper forms a comprehensive defense system covering both the execution and data layers, offering more thorough fixes and higher security.
\begin{table}[H]
  \caption{Validation of the Effectiveness of ContractTinker's Code Patches for the H-01.}
  \label{tab:h01_ct_patches}
  \centering
  \small
  \resizebox{\linewidth}{!}{%
  \begin{tabular}{p{5cm}>{\centering\arraybackslash}p{2cm}p{5.5cm}}
    \toprule
    \multicolumn{1}{c}{\textbf{Core Modification}} & \multicolumn{1}{c}{\textbf{Effective}} & \multicolumn{1}{c}{\textbf{Validity Analysis}} \\
    \midrule
    Adds the `nonReentrant' modifier to `crossSwap' & $\surd$ & Complies with official recommendations \\
    Swaps the order of fund transfers & $\times$ & Introduces a transaction failure bug \\
    \bottomrule
  \end{tabular}}
\end{table}
\begin{table}[htbp]
  \caption{Effectiveness of code patches generated by the proposed method against H-01.}
  \label{tab:h01_our_patches}
  \centering
  \small
  \resizebox{\linewidth}{!}{%
  \begin{tabular}{p{5cm}>{\centering\arraybackslash}p{2cm}p{5.5cm}}
    \toprule
    \multicolumn{1}{c}{\textbf{Core Modification}} & \multicolumn{1}{c}{\textbf{Valid}} & \multicolumn{1}{c}{\textbf{Validation Analysis}} \\
    \midrule
    Adds a mutual exclusion reentrancy lock to `crossSwap' & $\surd$ & Complies with official recommendations \\
    Uses `OracleLib.getAmountsOut' to calculate amounts & $\surd$ & Officially recommended alternative \\
    Adds a `verifyReserves' check before calculating the amount & Condition valid & Only trusted verification protects against fake reserves \\
    \bottomrule
  \end{tabular}}
\end{table}

\subsubsection{H-02 Vulnerability}

The principles and recommended fixes for the H-02 vulnerability have been explained in Section~3.1. This section focuses on analyzing the quality of the repair strategies generated by different tools and the actual effectiveness of the code patches. Table~\ref{tab:h02_strategies} presents the top three repair strategies proposed by ContractTinker and the method proposed in this paper. The results show that, among the three repair strategies proposed by ContractTinker, only the trading pair whitelist provides basic protection. The other two strategies, output amount verification and slippage protection, fail to address the root cause of the vulnerability. They cannot prevent attackers from stealing funds by constructing same-token transactions. Overall, ContractTinker's repair solution overlooks many critical issues. The repair strategies generated by the method proposed in this paper fully comply with the official repair requirements. Not only does it include a trusted transaction whitelist mechanism similar to ContractTinker, but it also takes the lead in implementing the officially required token non-identity verification. This verification directly blocks the core attack vector of the vulnerability. Furthermore, while the reserve verification strategy does not directly address this specific vulnerability, it serves as an additional layer of defense, enhancing the overall security of the protocol.

\begin{table}[htbp]
  \caption{Comparison of the Effectiveness of Different Repair Strategies for the H-02.}
  \label{tab:h02_strategies}
  \centering
  \small
  \resizebox{\linewidth}{!}{%
  \begin{tabular}{p{1.6cm}p{3cm}p{3cm}>{\centering\arraybackslash}p{1.2cm}p{3.5cm}}
    \toprule
    \multicolumn{1}{c}{\textbf{Method}} & \multicolumn{1}{c}{\textbf{Strategy}} & \multicolumn{1}{c}{\textbf{Description}} & \multicolumn{1}{c}{\textbf{Valid}} & \multicolumn{1}{c}{\textbf{Analysis}} \\
    \midrule
    \multirow{3}{*}{ContractTinker} & Trading pair whitelist & Ensure trading pairs are legitimate & $\surd$ & Officially recommended alternative \\
    & Output amount validation & Output amount greater than 0 & $\times$ & Does not address core vulnerability \\
    & Slippage protection & Add slippage limit & $\times$ & Does not address core vulnerability \\
    \midrule
    \multirow{3}{*}{Our Method} & Trading pair validation & Ensure tokens are different & $\surd$ & Complies with official recommendations \\
    & Trading pair whitelist & Ensure trading pairs are legitimate & $\surd$ & Officially recommended alternative \\
    & Reserve verification & Verify trading pair reserves & $\times$ & Does not address core vulnerability \\
    \bottomrule
  \end{tabular}}
\end{table}

Table~\ref{tab:h02_ct_patches} and Table~\ref{tab:h02_our_patches} present validity verification results for the code patches of the two types of fixes. The verification results show that neither of the two code patches proposed by ContractTinker addresses the root cause of the vulnerability. Both the validation of output amounts being greater than 0 and the consistency check of balances before and after withdrawal are merely superficial defensive measures. These methods cannot prevent attackers from core attack methods such as constructing same-token transactions and forging trading pairs. They completely fail to address the root cause of this vulnerability. The code patches in this paper directly comply with the official repair requirements. The verification of different transaction pairs directly implements the core check required by the official that `fromToken' is not equal to `toToken'. This check blocks the attack vector of same-token circular transactions at its source. The trading pair whitelist validation, through a trusted trading pair admission mechanism, completely prevents the possibility of forged malicious contracts participating in transactions. This strategy can serve as a valid alternative to the official recommendation.

\begin{table}[htbp]
  \caption{Validation of ContractTinker's Code Patch for the H-02 Vulnerability.}
  \label{tab:h02_ct_patches}
  \centering
  \small
  \resizebox{\linewidth}{!}{%
  \begin{tabular}{p{5cm}>{\centering\arraybackslash}p{2cm}p{5.5cm}}
    \toprule
    \multicolumn{1}{c}{\textbf{Core Modifications}} & \multicolumn{1}{c}{\textbf{Valid}} & \multicolumn{1}{c}{\textbf{Validity Analysis}} \\
    \midrule
    Validates that the output amount is greater than 0 & $\times$ & Does not address the core vulnerability \\
    Checks balance consistency before and after withdrawal & $\times$ & Does not address the core vulnerability \\
    \bottomrule
  \end{tabular}}
\end{table}

\begin{table}[htbp]
  \caption{Effectiveness of code patches generated by the proposed method against H-02.}
  \label{tab:h02_our_patches}
  \centering
  \small
  \resizebox{\linewidth}{!}{%
  \begin{tabular}{p{5cm}>{\centering\arraybackslash}p{2cm}p{5.5cm}}
    \toprule
    \multicolumn{1}{c}{\textbf{Core Modification}} & \multicolumn{1}{c}{\textbf{Valid}} & \multicolumn{1}{c}{\textbf{Validation Analysis}} \\
    \midrule
    Verifies that different trading pairs are used & $\surd$ & Complies with official recommendations \\
    Validates the trading pair whitelist & $\surd$ & Official alternative \\
    \bottomrule
  \end{tabular}}
\end{table}
\section{Conclusion}

Aiming at the core problems of existing smart contract vulnerability repair methods based on large language models, this paper combs through the smart contract security three-dimensional classification system and large model reasoning theory. Based on the tracking and analysis of real vulnerability cases, it reveals the shortcomings of incremental amplification of errors in linear thinking chains and the necessity of introducing a multi-branch reasoning mechanism. Furthermore, an automatic repair framework including structured analysis of audit reports, static analysis of programs based on Slither, a three-stage ToT reasoning framework, and a multi-level patch verification mechanism is designed and implemented. The framework uses self-consistent thought chains to improve the consistency of conclusions in the root cause analysis stage, and solves the stability problem of single-path reasoning through multi-branch exploration and state evaluation pruning in the repair strategy and patch generation stage. In this paper, comparative experiments and case analysis on 50 real high-risk vulnerabilities on the Code4Rena platform verify the effectiveness of the method. The results show that the method is superior to the baseline method, ContractTinker, in terms of the success rate of single and Top-3 repair strategies and the proportion of fully effective patches~\cite{sobania2023analysis, besta2024graph, koh2024tree}.

\section*{Acknowledgments}

AI-based tools were used for language polishing during manuscript preparation.

\bibliographystyle{unsrt}
\bibliography{references}

\end{document}